# Asymmetry approach to study for chemotherapy treatment and devices failure time's data using modified Power function distribution with some modified estimators

Azam Zaka[1,*], Ahmad Saeed Akhter[1] and Riffat Jabeen[2]

[1]College of Statistical and Actuarial Sciences, University of the Punjab, Lahore, Pakistan
[2]COMSATS University Islamabad Lahore Campus, Lahore, Pakistan

*Corresponding author: Tel: 92 300 4364368
E-mail: [1,*]azamzka@gmail.com (Azam Zaka);
[1]asakhter@yahoo.com (Ahmad S. Akhter);
[2]drriffatjabeen@cuilahore.edu.pk (Riffat Jabeen)

**Abstract**
In order to improve the already existing models that are used extensively in bio sciences and applied sciences research, a new class of Weighted Power function distribution (WPFD) has been proposed with its various properties and different modifications to be more applicable in real life. We have provided the mathematical derivations for the new distribution including moments, incomplete moments, conditional moments, inverse moments, mean residual function, vitality function, order statistics, mills ratio, information function, Shannon entropy, Bonferroni and Lorenz curves and quantile function. We have also characterized the WPFD, based on doubly truncated mean. The aim of the study is to increase the application of the Power function distribution. The main feature of the proposed distribution is that there is no induction of parameters as compare to the other generalization of the distributions, which are complexed having many parameters. We have used R programming to estimate the parameters of the new class of WPFD using Maximum Likelihood Method (MLM), Percentile Estimators (P.E) and their modified estimators. After analyzing the data, we conclude that the proposed model WPFD performs better in the data sets while compared to different competitor models.

**Biographical notes:**
Azam Zaka is a PhD student in the College of Statistical and Actuarial Sciences, University of the Punjab, Lahore, Pakistan. He gained his MPhil and MSc in Statistics from the College of Statistical and Actuarial Sciences, University of the Punjab, Lahore, Pakistan. He is presently working as Assistant Professor of Statistics at the Govt. College of Science, Wahdat Road, Lahore, Pakistan. His research interests are distribution theory, statistical inference and quality control charts.

Dr. Ahmad Saeed Akhter is working as a Professor of Statistics at College of Statistical and Actuarial Sciences, University of the Punjab, Lahore. He has 37 years of teaching experience at university level. He has several research publications to his credit in different reputed national and international conferences and journals. His areas of interests are distribution theory and statistical inference.

Dr. Riffat Jabeen has earned a doctoral degree in Statistics from National College of Business Administration and Economics Lahore Pakistan. She has an M.Phil degree in Statistics from College of Statistical and Actuarial Sciences, University of the Punjab, Lahore, Pakistan, and Master in Statistics from LCWU, Lahore, Pakistan. Dr. Riffat Jabeen is currently serving for COMSATS University Islamabad, Lahore Campus, as an Assistant Professor at Department of Statistics. Her research interests are distribution theory, quality control charts and their applications, and survey sampling.

## 1. Introduction

Weighted distributions have been extensively applied in the field of sampling which deals with unequal weighting of the units for example actuarial sciences, biomedicine, ecology and survival data analysis. Fisher (1934) has used firstly weighted distributions in order to estimate the frequencies by using methods of ascertainment.

Let we have a random variable x with the following probability density function,

$$f(x; \alpha, \beta) = \frac{w(x;\beta)f(x;\alpha)}{E[w(x;\beta)]} \qquad (1.1)$$

We take $w(x; \beta)$ as the non-negative weight function.

Patil and Ord (1976) utilized the concept of weighted distribution and presented the idea of $\beta th$ order size biased distribution utilizing the weight function as $w(x) = x^\beta$, and that was called moment distribution. It is is called as size biased when $\beta=1$, whereas it is called the area biased distribution for $\beta =2$. Afterwards many statisticians worked on weighted distribution such as Patil and Rao (1978), Arnold and Nagaraja (1991), Gove (2003), Mir and Ahmed (2009), Das and Roy (2011) applied this concept on different probability distributions. Ramos and Louszada (2016) discussed generalized weighted Lindley distributions with its different properties. Dar et al. (2017) introduced transmuted weighted exponential distribution and discussed its application. Balakrishnan et al. (2017) introduced the weighted Poisson distribution and its application to cure rate models. Different works on the weighted distributions and its parameters estimations are discussed in (Para and Jan (2018), Perveen and Ahmad (2018), Acitas (2019))

Dallas (1976) introduced the power function as the inverse of Pareto distribution. Meniconi and Barry (1996) showed that Power function distribution (PFD) is better to fit for failure data over exponential, lognormal and Weibull because it provides a better fit. Zaka and Akhter (2013) worked on parameters estimation for Power function distribution. Afterwards Zaka and Akhter (2014) provided the different modifications and Bayes inference of the parameters from power function distribution. Zaka et al. (2020) proposed the exponentiated class of Power function distribution.

In this research paper, the effort is to introduce a new model called the Weighted Power function distribution (WPFD) which may be more suitable to the applied bio sciences and applied sciences data. We have studied the various properties of the under discussion distribution as moments, inverse moments, conditional moments, moments generating function, quantile function, mean residual function, vitality function, information function, mills ratio, bonferroni curve, lorenz curve, some entropies and order statistics. We have also produced some modifications of the WPFD. We have demonstrated the performance of the new models over already existing distributions by using a real life example from medical and applied sciences. The main feature of the proposed distributions is that there is no induction of parameters as compare to the other generalization of the distributions, which are complexed having many parameters.

## 2. Weighted Power Function Distribution (WPFD)

Power function distribution (PFD) may model life time data as a good fit. The pdf (probability distribution function) may be written as:

$$f(x) = \frac{\gamma x^{\gamma-1}}{\beta^\gamma} \; ; \qquad 0 < x < \beta \tag{2.1}$$

and 
$$F(x) = \left(\frac{x}{\beta}\right)^\gamma \; ; \text{ where } \beta \text{ and } \gamma \text{ are the scale and shape parameters.}$$

We may consider the following weight function as:
$$w(x; \alpha) = F(\alpha x) \tag{2.2}$$

Hence using (2.1) and (2.2) in (1.1), the pdf of the Weighted Power function distribution (WPFD) is

$$g(x) = \frac{2\gamma\, x^{2\gamma-1}}{\beta^{2\gamma}} \; ; \qquad 0 < x < \beta \tag{2.3}$$

The cumulative distribution function (cdf), survival, and hazard functions of WPFD are

$$G(x) = \left(\frac{x}{\beta}\right)^{2\gamma} \tag{2.4}$$

$$s(x) = 1 - \left(\frac{x}{\beta}\right)^{2\gamma} \tag{2.5}$$

$$h(x) = \frac{(2\gamma)(X)^{2\gamma-1}}{(\beta)^{2\gamma} - (x)^{2\gamma}} \tag{2.6}$$

### 3. Asymptotic Behavior

We may see the asymptotic behavior of the pdf, cdf, hazard and survival functions of WPFD as $x \to 0$ and $x \to \infty$.

i. $lim_{x \to 0} g(x) = 0$ ; if $\gamma = 0$ and $\beta > 0$.
ii. $lim_{x \to \infty} g(x) = \infty$ ; $\forall$ possible values of $\gamma$ and $\beta$.
iii. $lim_{x \to 0} G(x) = 0$ ; $\forall$ possible values of $\gamma$ and $\beta$.
iv. $lim_{x \to 0} G(x) = 1$ ; if $\gamma = 0$ and $\beta > 0$.
v. $lim_{x \to 0} S(x) = 0$ ; if $\gamma > 0$ and $\beta > 0$.
vi. $lim_{x \to 0} S(x) = \infty$ ; if $\gamma > 0$ and $\beta = 0$.
vii. $lim_{x \to 0} S(x) = 1$ ; if $\gamma = 0$ and $\beta \geq 1$.
viii. $lim_{x \to \infty} S(x) = \infty$ ; if $\gamma > 0$ and $\beta \geq 1$.
ix. $lim_{x \to 0} h(x) = 0$ ; if $\gamma = 0$ and $\beta > 0$.
x. $lim_{x \to \infty} h(x) = 0$ ; if $\beta > 0$ and $\gamma \geq 1$.

*3.1. Characteristics of Hazard function using Glaser method*

We may use the conditions defined by Glaser (1980) as

$$\eta(x) = -\frac{g'(x)}{g(x)}$$

$$\eta(x) = -\frac{(2\gamma - 1)}{x}$$

$$\acute{\eta}(x) = \frac{(2\gamma - 1)}{x^2}$$

If $x > 0$, then $\acute{\eta}(x) > 0$ under the following conditions
   i. If $\gamma \geq 1$, then $\acute{\eta}(x) > 0$.
   ii. If $\gamma = 0$, then $\acute{\eta}(x) = 0$.

iii. If $\gamma < 1$ or $\gamma = 0$, then $\acute{\eta}(x) < 0$

The above conditions shows that the hazard function of WPFD is increasing but if $\gamma < 1$ or $\gamma = 0$, then it will be decreasing function.

*3.2. Shapes*

Figures.1-3 (See Appendix) shows some plots of the pdf, cdf and hrf for some parameter values of WPFD. WPFD have different shapes like increasing, right and left skewed and J shapes.

## 4. Mathematical Properties of the WPFD

We may discuss some general properties of WPFD under this current section;

*4.1. Quantile Function*

By inverting (2.4), we get the quantile function as:

$$Q(u) = \beta U^{(1/2\gamma)}$$

*4.2. Moments and Inverse Moments*

The $r^{th}$ moment and the $r^{th}$ inverse moment of the random variable "X", say $\mu'_r$ may be expressed as

$$\mu'_r = \frac{2\gamma \beta^r}{(r+2\gamma)} \text{ and } \mu'_{(-r)} = \frac{2\gamma \beta^{-r}}{(-r+2\gamma)}$$

*4.3. Incomplete Moments and Conditional Moments*

The Incomplete and Conditional moments may be expressed as:

$$\mu_{x|(\beta,\gamma);r}(p) = \int_0^p x^r \frac{2\gamma x^{2\gamma-1}}{\beta^{2\gamma}} dx = \left(\frac{2\gamma}{\beta^{2\gamma}}\right) \frac{(p)^{r+2\gamma}}{r+2\gamma}$$

And Conditional moments of "X"

$$E(x^r | x > t) = \left(\frac{2\gamma}{\bar{F}(t)\beta^{2\gamma}}\right) \frac{(\beta)^{r+2\gamma} - (t)^{r+2\gamma}}{r+2\gamma}$$

*4.4. Moments Generating Function (MGF)*

The MGF of WPFD is expressed as

$$M_0(t) = 1 + \sum_{r=1}^{\infty} \frac{(t\beta)^r}{r!\left(\frac{r}{2\gamma}+1\right)}$$

*4.5. Mean Residual function(MRF) and Vitality function(VF)*

The mean residual function is given by the relation:

$$e(x) = \frac{\int_x^\infty S(t)dt}{S(x)} = \frac{(\beta-x) - \frac{1}{\beta^{2\gamma}}\left(\frac{\beta^{2\gamma+1} - x^{2\gamma+1}}{2\gamma+1}\right)}{1-\left(\frac{x}{\beta}\right)^{2\gamma}}$$

and

$$V(x) = \frac{\int_x^\infty x f(x) dx}{S(x)} = \frac{\frac{2\gamma}{\beta^{2\gamma}}\left(\frac{\beta^{2\gamma+1} - x^{2\gamma+1}}{2\gamma+1}\right)}{1-\left(\frac{x}{\beta}\right)^{2\gamma}}$$

*4.6. Some Entropies and Information function*

The Reńyi entropy of a random variable "X" is defined as;

$$I_R(s) = \frac{1}{1-s} \log\left[\int_0^\infty f^s(x) dx\right] = \frac{1}{1-s} \log\left\{\left(\frac{2\gamma}{\beta^{2\gamma}}\right)^s \left(\frac{\beta^{s(2\gamma-1)+1}}{s(2\gamma-1)+1}\right)\right\}$$

And Shannon entropy of "X" is defined as

$$E\{-\log[f(x)]\} = -\left[\log\frac{2\gamma}{\beta^{2\gamma}} + (2\gamma - 1)\left\{\log\beta - \frac{1}{2\gamma}\right\}\right]$$

Also Information function provides the moments of self-information of the probability density function by taking the derivatives at certain at certain place

$$E\{f(x)\}^s = \left(\frac{2\gamma}{\beta^{2\gamma}}\right)^s \left(\frac{\beta^{s(2\gamma-1)+1}}{s(2\gamma-1)+1}\right)$$

*4.7. Order Statistics*

The pdf of *jth* order statistics may be written as following,

$$f_{j,n}(x) = \frac{n!}{(j-1)!(n-i)!} f(x) F^{j-1}(x)\{1-F(x)\}^{n-j}; \quad j = 1, \ldots, n$$

Therefore the pdf of lower order statistics

$$f_{1,n}(x) = \left\{\frac{n(2\gamma)(X)^{2\gamma-1}}{\beta^{2\gamma}}\right\} \left(1 - \frac{(x)^{2\gamma}}{\beta^{2\gamma}}\right)^{n-1}$$

And the pdf of highest order statistic

$$f_{n,n}(x) = \left\{\frac{n(2\gamma)(X)^{2\gamma-1}}{\beta^{2\gamma}}\right\} \left(\frac{(x)^{2\gamma}}{\beta^{2\gamma}}\right)^{n-1}$$

*4.8. The Mills Ratio*

This is defined as the inverse of hazard rate function and mathematically expressed as

$$m(x) = \frac{s(x)}{g(x)} = \frac{(\beta^{2\gamma} - x^{2\gamma})}{(2\gamma)x^{2\gamma-1}}$$

*4.9. Bonferroni and Lorenz curves*

$$L(p) = \frac{1}{\mu} \int_0^q x\, f(x) dx = \frac{(2\gamma)(q)^{2\gamma+1}}{\mu\, \beta^{2\gamma}\, (2\gamma+1)}$$

and
$$B(p) = L(p)/p$$

## 5. Some Modified Weighted Power Function Distribution (MWPFD)

*5.1. 1$^{st}$ Modified Weighted Power Function Distribution (MWPFD-1)*

In this modification we replace the weight function in (2.2) by $F(\alpha^\theta x^\theta)$ i.e.

$w(x;\alpha,\theta) = F(\alpha^\theta x^\theta)$  (See Table 1 in Appendix)

*5.2. 2$^{nd}$ Modified Weighted Power Function Distribution (MWPFD-2)*

In this modification we replace the weight function in (2.2) by $F(\alpha^{\frac{1}{\theta}} x^{\frac{1}{\theta}})$ i.e.

$w(x;\alpha,\theta) = F(\alpha^{\frac{1}{\theta}} x^{\frac{1}{\theta}})$   (See Table 1 in Appendix)

*5.3. Characterization based on Conditional moment (Doubly Truncated Mean)*

Let *"X"* Weighted Power function Variable with Probability density function

$$g(x) = \frac{2\gamma\, x^{2\gamma-1}}{\beta^{2\gamma}}\,; \qquad 0 < x < \beta$$

And let $\bar{G}(x)$ be the survival function respectively. Then the random variable *"X"* has Weighted Power function distribution if and only if

$$E(X|x < X < y) = \frac{2\gamma}{\beta^{2\gamma}\{G(y)-G(x)\}} \left[\frac{y^{2\gamma+1} - x^{2\gamma+1}}{2\gamma+1}\right]$$

where $E(X|x \leq X \leq y)$: Doubly Truncated Mean.

Proof:

Necessary part:

$$E(X|x \leq X \leq y) = \frac{1}{G(y)-G(x)} \int_x^y x \frac{2\gamma\, x^{2\gamma-1}}{\beta^{2\gamma}} dx$$

$$E(X|x < X < y) = \frac{2\gamma}{\beta^{2\gamma-1}\{G(y)-G(x)\}} \left[\frac{y^{2\gamma+1} - x^{2\gamma+1}}{2\gamma+1}\right] \qquad (5.1)$$

Now Sufficient Part:

$$E(X|x \leq X \leq y) = \frac{1}{\{G(y)-G(x)\}} \int_x^y x\, g(x) dx$$

$$E(X|x \leq X \leq y) = \frac{yG(y) - xG(x) - \int_x^y G(X)\, dx}{G(y)-G(x)} \qquad (5.2)$$

Equate (5.1) and (5.2), we get

$$\frac{yG(y) - xG(x) - \int_x^y G(x)\, dx}{G(y)-G(x)} = \frac{2\gamma}{\beta^{2\gamma}\{G(y)-G(x)\}} \left[\frac{y^{2\gamma+1} - x^{2\gamma+1}}{2\gamma+1}\right]$$

After differentiating the above equation, we get

$$g(y) = \frac{2\gamma\, y^{2\gamma-1}}{\beta^{2\gamma}}$$

This is the pdf of WPFD.

## 6. Comparison between Maximum Likelihood and Percentile Estimation Methods of the Parameters of WPFD

*6.1. Maximum Likelihood Method (MLM)*

Let $x_1, x_2, \ldots, x_n$ be a random sample of size n from the WPFD. The log-likelihood function for the WPFD is given by

$$L(\gamma, \beta) = n\ln(2\gamma) + (2\gamma - 1) \sum_{i=1}^{n} \ln(x_i) - 2n\gamma \ln(\beta)$$

The score vector is

$$U_\beta(\gamma, \beta) = \frac{n\gamma}{\beta} \quad (6.1)$$

$$U_\gamma(\gamma, \beta) = \frac{n}{\gamma} + 2\sum_{i=1}^{n} \ln x_i - 2n \ln(\beta) \quad (6.2)$$

The parameters of Weighted Power Function distribution can be obtained by solving the above equations resulting from setting the two partial derivatives of $L(\gamma, \beta)$ to zero;

$\beta$ does not exist, but the likelihood function can be maximize by taking

$$\hat{\beta} = x_n \; ; \quad \text{Where "} x_n \text{" is the maximum value in the data.} \quad (6.3)$$

$$\hat{\gamma} = \left( \frac{n}{2(n \ln(\beta) - \sum_{i=1}^{n} \ln x_i)} \right)$$

*6.2. Modified Maximum Likelihood Method (MMLM)*

In this modification of the MLM, the (6.2) equation is replaced by the co-efficient of variation of WPFD.

$$c.v = \frac{1}{\sqrt{4\gamma(\gamma + 2)}}$$

By solving the above expression, we get

$$\hat{\gamma} = \frac{-1 + \sqrt{1 + \frac{\bar{x}^2}{s^2}}}{2}$$

$\hat{\beta} = x_n \; ;$ Where "$x_n$" is the maximum value in the data

*6.3. Estimation of Weighted Power Function Distribution Parameters from "common percentiles" (P.E)*

Dubey (1967) proposed a percentile estimator of the shape parameter, based on any two sample percentiles. Marks (2005) also discussed it, in which he estimated the parameters of Weibull distribution with the help of percentiles. Let $x_1, x_2, x_3, \ldots, x_n$ be a random sample of size n drawn from Probability density function of Weighted Power function distribution. The cumulative distribution function of a Weighted Power function distribution with shape and scale parameters $\beta$ and $\gamma$, respectively

$$x = \beta(R)^{1/2\gamma} \; ; \quad R = G(x) \quad (6.4)$$

Let $P_{75}$ and $P_{25}$ are the 75<sup>th</sup> and 25<sup>th</sup> Percentiles, therefore (6.4) becomes

$$P_{75} = \beta(.75)^{1/2\gamma} \quad (6.5)$$

$$P_{25} = \beta(.25)^{1/2\gamma} \quad (6.6)$$

Solving the above equations, we get

$$\hat{\gamma} = \frac{\ln\left(\frac{.75}{.25}\right)}{2 * \ln\left(\frac{P_{75}}{P_{25}}\right)} \quad \text{and} \quad \hat{\beta} = \frac{P_{75}}{(.75)^{1/2\hat{\gamma}}}$$

generally $\quad \hat{\gamma} = \dfrac{\ln\left(\frac{H}{L}\right)}{2 * \ln\left(\frac{P_H}{P_L}\right)} \quad$ and $\quad \hat{\beta} = \dfrac{P_H}{(H)^{1/2\hat{\gamma}}}$

Where H= Maximum Percentage, L= Minimum Percentage and P = Percentile

*6.4. Modified Percentile Estimator (M.P.E)*

In this modification of the percentile estimators, (6.6) is replaced by the Median of Weighted Power function distribution.

$$\tilde{x} = \frac{\beta}{2^{1/2*\gamma}} \quad \Rightarrow \quad \hat{\beta} = \tilde{x} 2^{1/2\gamma}$$

From (6.5) $\quad \hat{\beta} = \dfrac{P_{75}}{(.75)^{1/2\gamma}}$

therefore
$$\tilde{x}2^{1/2\gamma} = \frac{P_{75}}{(.75)^{1/2\gamma}} \Rightarrow \hat{\gamma} = \frac{\ln(2*.75)}{2*\ln(\frac{P_{75}}{\tilde{x}})}$$

$$\hat{\gamma} = \frac{\ln(2*H)}{2*\ln(\frac{P_H}{\tilde{x}})} \text{ and } \hat{\beta} = \frac{P_H}{(H)^{1/2\gamma}}$$

Where H= Maximum Percentage and P = Percentile.

A simulation study is used in order to compare the performance of the proposed estimation methods. We carry out this comparison taking the samples of sizes as n = 40 and 100 with pairs of (β, γ) = {(1, 2), (3, 2) and (4, 3)}. We generated random samples of different sizes by observing that if $R_i$ is random number taking (0, 1), then $x_i = \beta R_i^{1/2\gamma}$ is the random number generation from Weighted Power function distribution with (γ, β) parameters. All results are based on 5000 replications. Such generated data have been used to obtain estimates of the unknown parameters. The results obtained from parameters estimation of the 2-parameters of Weighted Power function distribution using different sample sizes and different values of parameters with mean square error M.S.E.

$$M.S.E\,(\hat{\beta}) = E\left[(\hat{\beta} - \beta)^2\right], M.S.E\,(\hat{\gamma}) = E[(\hat{\gamma} - \gamma)^2]$$

If we study the two results of the Table 2 and Table 3 from Appendix, in which sample sizes are (40 and 100) and the combinations of the values of (β, γ) = {(1, 2), (3, 2) and (4, 3)}. Then we get the results that MLM is the best for the estimation of β and γ. After MLM, the MMLM and Percentile method are best for the estimation of scale and shape parameters of the Weighted Power function distribution.

7. **Application**

In this section, we illustrate the usefulness of the WPFD and its modifications. We fit these distributions on real life data and compare the result with the existing distributions.

*7.1. The Data about group of patients given Chemotherapy treatment*

The first data set is reported by Bekker et al. (2000), which corresponds to the survival times (in years) of a group of patients given chemotherapy treatment alone. The data consisting of survival times (in years) for 46 patients are:
0.047,0.115, 0.121,0.132,0.164,0.197,0.203,0.260,0.282,0.296, 0.334, 0.395, 0.458, 0.466, 0.501, 0.507, 0.529, 0.534, 0.540, 0.641, 0.644, 0.696, 0.841, 0.863, 1.099, 1.219, 1.271, 1.326, 1.447, 1.485, 1.553, 1.581, 1.589, 2.178, 2.343, 2.416, 2.444, 2.825, 2.830, 3.578, 3.658, 3.743, 3.978, 4.003, 4.033. We have estimated the parameters of the model by the method of MLE. We have used five other criteria's to compare the performance of the proposed distributions with already existing distributions. We have used Akaike information criterion (AIC), consistent Akaike information criterion (CAIC), Bayesian information criterion (BIC) and Hannan-Quinn information criterion (HQIC) for this comparison.

We have compared our proposed distribution with the kumarswamy Marshal-Olkin family of distribution (Kw-MO) proposed by Alizadeh et al. (2015), Kumaraswamy Power function distribution (KPFD) by Ibrahim (2017), McDonald`s Power function distribution (McPFD) by Haq et al. (2018) and Power function distribution (PFD) for the same data set. The TTT-plot is displayed in Figure 4 (See Appendix), which indicates that the HRF associated with the data set has a bathtub shape, since the plot shows a first concave curvature. So, we can easily fit WPFD on the Chemotherapy treatment data. In Table 4 (See Appendix), we may see that WPFD provides better fit for the above data set as it provides minimum AIC, BIC, CAIC and HQIC.

*7.2. Devices failure times data*

The second data set refers to 30 devices failure times given in Table 15.1 by Meeker and Escobar (1998). The data are: 275, 13, 147, 23, 181, 30, 65, 10, 300, 173, 106, 300, 300, 212, 300, 300, 300, 2, 261, 293, 88,247, 28, 143, 300, 23, 300, 80, 245, and 266. The same data has been used by Tahir et al. (2016) for Weibull Power function distribution (WP). We have used this data set to show the performance of our proposed Weighted Power function distribution (WPFD) over Tahir et al. (2016), Kumaraswamy Power function distribution (KPFD) by Ibrahim (2017), McDonald`s Power function distribution (McPFD) by Haq et al. (2018) and Power function distribution (PFD).

The TTT-plot is displayed in Figure 5 (See Appendix), which indicates that the HRF associated with the data set has an increasing shape, since the plot shows a first concave curvature. So, we can easily fit WPFD on the Devices failure time's data. In Table 5 (See Appendix) provide the Statistics for Devices failure times. The proposed model WPFD is showing better results by providing the smallest AIC, BIC, CAIC and HQIC for the devices failure time's data.

8. **Concluding Remarks**

We have seen from Table 4 and 5 that WPFD best describe the discussed data sets as compare to the other models in literature. So may be used in order to describe the nature of life time data of medical sciences and applied sciences by providing the lowest values of the AIC, BIC, CAIC and HQIC among all fitted probability distribution functions.

In this paper, we proposed the WPFD and its modifications. We derived some of its properties. The parameters of the distribution have been estimated by the Maximum Likelihood Method (MLM), Percentile Estimators (P.E) and their modified estimators. We have also characterized the distribution by doubly truncated mean (DTM). Different criteria's has been used as discussed above to prove that the WPFD provides a better fit than existing distributions. It is hoped that the findings of this paper will be useful for researchers in different field of applied sciences.

# Appendix

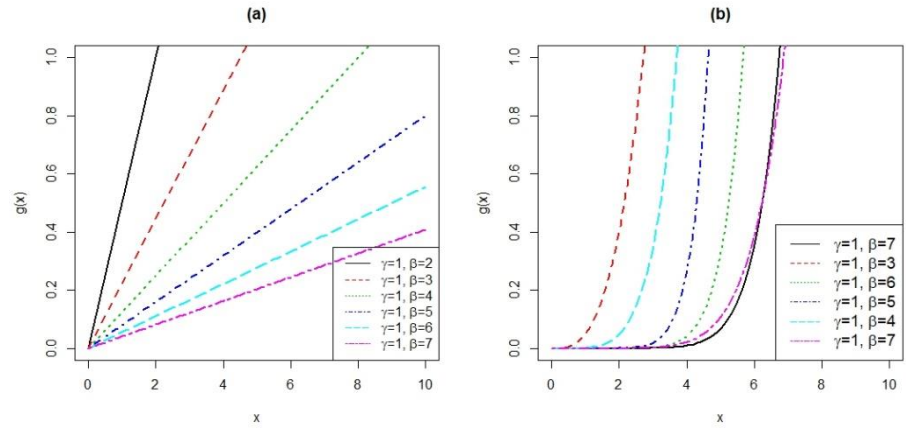

**Figure 1:** Plots of pdf of WPFD.

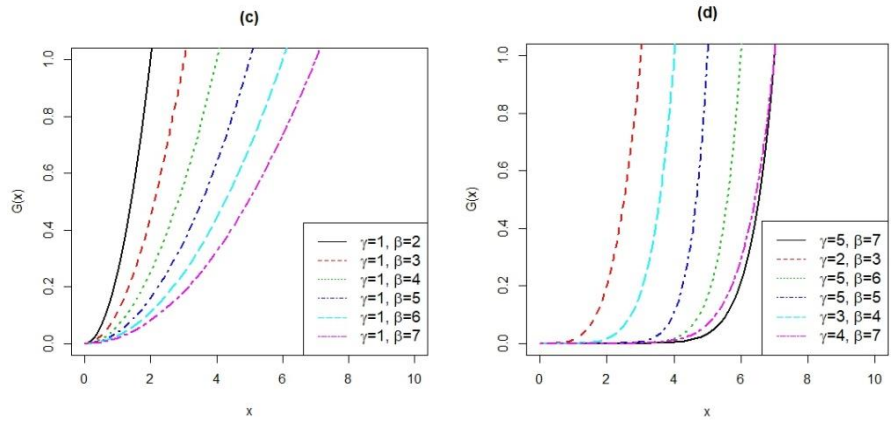

**Figure 2:** Plots of cdf of WPFD.

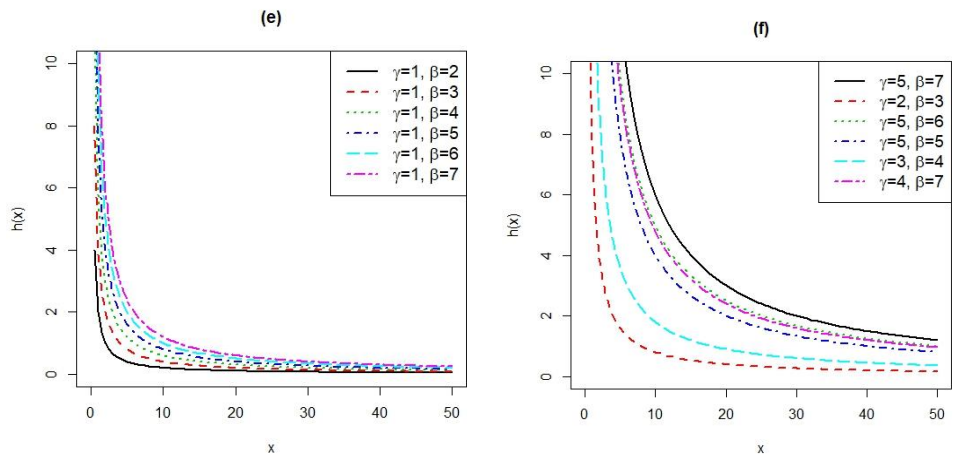

**Figure 3:** Plots of hrf of WPFD.

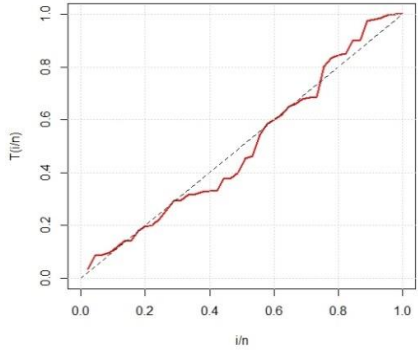 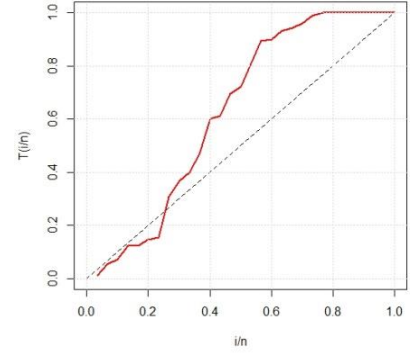

Figure 4: TTT Plot for Chemotherapy treatment    Figure 5: TTT Plot for Devices failure times

Table 1: The pdf and properties of the MWPFD-1 and MWPFD-2

| SR | Properties | MWPFD-1 | MWPFD-2 |
|---|---|---|---|
| 1 | Complete pdf | $\dfrac{(\gamma\theta+\gamma)(x)^{\gamma\theta+\gamma-1}}{\beta^{\gamma\theta+\gamma-1}}; \ 0<x<\beta$ where $\beta>0$ and $\theta \geq 0$ | $\dfrac{(\gamma/\theta+\gamma)(x)^{\gamma/\theta+\gamma-1}}{\beta^{\gamma/\theta+\gamma}}; \ 0<x<\beta$ where $\beta>0$ and $\theta>0$ |
| 2 | Moments | $\dfrac{(\gamma\theta+\gamma)\beta^r}{(r+(\gamma\theta+\gamma))}$ | $\dfrac{(\gamma/\theta+\gamma)\beta^r}{(r+(\gamma/\theta+\gamma))}$ |
| 3 | cdf | $\dfrac{(x)^{(\gamma\theta+\gamma)}}{\beta^{(\gamma\theta+\gamma)}}$ | $\dfrac{(x)^{(\gamma/\theta+\gamma)}}{\beta^{(\gamma/\theta+\gamma)}}$ |
| 4 | Moments Generating Function | $1 + \sum_{r=1}^{\infty} \dfrac{(t\beta)^r}{r!\left(\dfrac{r}{(\gamma\theta+\gamma)}+1\right)}$ | $1 + \sum_{r=1}^{\infty} \dfrac{(t\beta)^r}{r!\left(\dfrac{r}{(\gamma/\theta+\gamma)}+1\right)}$ |
| 5 | Survival Function | $1 - \dfrac{(x)^{(\gamma\theta+\gamma)}}{\beta^{(\gamma\theta+\gamma)}}$ | $1 - \dfrac{(x)^{(\gamma/\theta+\gamma)}}{\beta^{(\gamma/\theta+\gamma)}}$ |
| 6 | Hazard Function | $\dfrac{(\gamma\theta+\gamma)(X)^{(\gamma\theta+\gamma)-1}}{(\beta)^{(\gamma\theta+\gamma)}-(x)^{(\gamma\theta+\gamma)}}$ | $\dfrac{(\gamma/\theta+\gamma)(X)^{(\gamma/\beta+\gamma)-1}}{(\beta)^{(\gamma/\theta+\gamma)}-(x)^{(\gamma/\theta+\gamma)}}$ |
| 7 | Random Number Generator | $\beta(R)^{\frac{1}{(\gamma\theta+\gamma)}}$ | $\beta(R)^{\frac{1}{(\gamma/\theta+\gamma)}}$ |
| 8 | Inverse Moments | $\dfrac{(\gamma\theta+\gamma)\beta^{-r}}{(-r+(\gamma\theta+\gamma))}$ | $\dfrac{(\gamma/\theta+\gamma)\beta^{-r}}{(-r+(\gamma/\theta+\gamma))}$ |

Table 2: Estimates for the parameters of Weighted Power function distribution with different estimation methods under the sample size 40

| Methods | True Values | | Estimated Values | | M.S.E | |
|---|---|---|---|---|---|---|
| | $\beta$ | $\gamma$ | $\hat{\beta}$ | $\hat{\gamma}$ | $\hat{\beta}$ | $\hat{\gamma}$ |
| MLM | 1 | 2 | 0.9938524 | 2.104826 | 0.00007404 | 0.1308768 |
| | 3 | 2 | 2.981884 | 2.098744 | 0.0006498764 | 0.1283807 |
| | 4 | 3 | 3.983295 | 3.154041 | 0.0005503625 | 0.2938336 |
| MMLM | 1 | 2 | 0.9936337 | 2.079167 | 0.00008128 | 0.1669906 |
| | 3 | 2 | 2.981997 | 2.072701 | 0.0006456896 | 0.1585582 |
| | 4 | 3 | 3.9833 | 3.146969 | 0.0005463373 | 0.4190948 |
| P.E | 1 | 2 | 0.9927462 | 2.177096 | 0.0008329366 | 0.3118459 |
| | 3 | 2 | 2.97946 | 2.181738 | 0.007721837 | 0.3214192 |
| | 4 | 3 | 3.980408 | 3.267518 | 0.006049237 | 0.6997704 |
| M.P.E | 1 | 2 | 0.9942221 | 2.249254 | 0.001039149 | 0.6753827 |
| | 3 | 2 | 2.980275 | 2.273176 | 0.009635285 | 0.7130634 |

| | 4 | 3 | 3.982908 | 3.418103 | 0.007461182 | 1.716589 |

**Table 3: Estimates for the parameters of Weighted Power function distribution with different estimation methods under the sample size 100**

| Methods | True Values | | Estimated Values | | M.S.E | |
|---|---|---|---|---|---|---|
| | $\beta$ | $\gamma$ | $\hat{\beta}$ | $\hat{\gamma}$ | $\hat{\beta}$ | $\hat{\gamma}$ |
| MLM | 1 | 2 | 0.9974972 | 2.039611 | 0.00001267 | 0.0452979 |
| | 3 | 2 | 2.99268 | 2.041316 | 0.0001078838 | 0.04401944 |
| | 4 | 3 | 3.993276 | 3.05943 | 0.00009112 | 0.1004658 |
| MLM-1 | 1 | 2 | 0.9974775 | 2.03012 | 0.00001305 | 0.05853799 |
| | 3 | 2 | 2.992459 | 2.026962 | 0.000112158 | 0.05775556 |
| | 4 | 3 | 3.993253 | 3.046534 | 0.00009274 | 0.1397898 |
| P.E | 1 | 2 | 0.9976476 | 2.064303 | 0.0003183863 | 0.09982225 |
| | 3 | 2 | 2.992453 | 2.064008 | 0.003019171 | 0.09917686 |
| | 4 | 3 | 3.992342 | 3.117756 | 0.002340664 | 0.2392223 |
| P.E-1 | 1 | 2 | 0.9975793 | 2.101732 | 0.0004208572 | 0.1971112 |
| | 3 | 2 | 2.992632 | 2.110028 | 0.003701643 | 0.2164466 |
| | 4 | 3 | 3.993442 | 3.144727 | 0.002941814 | 0.4740981 |

**Table 4: Statistics for Chemotherapy Treatment Data**

| Distribution | AIC | CAIC | BIC | HQIC |
|---|---|---|---|---|
| **WPFD** | **107.5513** | **107.6489** | **109.3125** | **108.2008** |
| PFD | 108.7513 | 108.7482 | 110.5125 | 109.3788 |
| MWPFD-1 | 109.5513 | 109.8513 | 113.0737 | 110.8503 |
| MWPFD-2 | 109.6419 | 109.8711 | 113.0934 | 110.8704 |
| KPFD | 109.8881 | 109.9035 | 114.4717 | 111.1366 |
| McPFD | 111.1315 | 112.1841 | 118.1763 | 113.7294 |
| Kw-MOW | 119.134 | 120.672 | 128.167 | 122.501 |

**Table 5: Statistics for Devices failure times**

| Distribution | AIC | CAIC | BIC | HQIC |
|---|---|---|---|---|
| **WPFD** | **250.5577** | **250.7577** | **251.6487** | **250.8147** |
| MWPFD-1 | 252.5577 | 253.1893 | 254.7398 | 253.0717 |
| MWPFD-2 | 252.7557 | 253.3894 | 254.9378 | 253.5713 |
| KPFD | 254.2257 | 255.559 | 257.4988 | 254.9968 |
| McPFD | 256.2535 | 258.6064 | 260.6177 | 257.2816 |
| PFD | 258.5577 | 258.7577 | 261.6487 | 259.8147 |
| WP | 311.1535 | 312.0766 | 315.3571 | 312.4983 |